\documentclass[prl, onecolumn,showpacs,10pt, preprint]{revtex4-1}

\usepackage{amsmath}
\usepackage{amssymb}
\usepackage[final]{graphicx}
\usepackage{upgreek}
\usepackage{ulem}
\usepackage[final]{pdfpages}

\newcommand{\mueV}{\,\upmu\mathrm{eV}}

\begin{document}

\title{Driven-dissipative confined polariton condensate under magnetic field}
\author{C. Sturm}\email[]{present address: Institut f\"ur Experimentelle Physik II, Universit\"at Leipzig, Linnestr. 5, 04103 Leipzig, Germany}
\affiliation{Laboratoire de Photonique et Nanostructures, LPN/CNRS, Route de Nozay, 91460 Marcoussis, France}
 \author{D. Solnyshkov}
 \affiliation{Institut Pascal, Photon-N2, Clermont Universit\'{e} and Universit\'{e} Blaise Pascal, CNRS, 63177 Aubi\`{e}re Cedex, France}
  \author{O. Krebs}
 \affiliation{Laboratoire de Photonique et Nanostructures, LPN/CNRS, Route de Nozay, 91460 Marcoussis, France}
  \author{A.~Lema\^{\i}tre}
 \affiliation{Laboratoire de Photonique et Nanostructures, LPN/CNRS, Route de Nozay, 91460 Marcoussis, France}
  \author{I.~Sagnes}
 \affiliation{Laboratoire de Photonique et Nanostructures, LPN/CNRS, Route de Nozay, 91460 Marcoussis, France}
  \author{E.~Galopin}
 \affiliation{Laboratoire de Photonique et Nanostructures, LPN/CNRS, Route de Nozay, 91460 Marcoussis, France}
  \author{A. Amo}
 \affiliation{Laboratoire de Photonique et Nanostructures, LPN/CNRS, Route de Nozay, 91460 Marcoussis, France}
  \author{G. Malpuech}
 \affiliation{Institut Pascal, Photon-N2, Clermont Universit\'{e} and Universit\'{e} Blaise Pascal, CNRS, 63177 Aubi\`{e}re Cedex, France}
 \author{J. Bloch}
\affiliation{Laboratoire de Photonique et Nanostructures, LPN/CNRS, Route de Nozay, 91460 Marcoussis, France}
\date{\today}

\begin{abstract}
We investigate exciton-polariton condensation under magnetic field in a single high-quality semiconductor micropillar cavity. We observe successive polariton condensation of  each spin component for two distinct threshold powers. Pronounced and non-monotonous variations of both the Zeeman splitting and the circular polarization of the emission are measured across these two condensation thresholds. This unexpected behavior deeply deviates from the so-called spin Meissner effect predicted for a fully thermalized system. Our measurements can be understood in a kinetic approach taking into account spin-anisotropic interactions within the entire system: the polariton condensate and the cloud of uncondensed excitons.



\end{abstract}


\maketitle

Manipulation of the laser emission using magnetic fields has remained elusive due to the weak magnetic response of photonic devices, particularly in the visible range. In the case of semiconductor VCSELs, intrinsic birefringence due to anisotropies along the main crystal axes greatly reduces the influence of Zeeman or Faraday effects on the polarisation of the lasing modes~\cite{vanDoorn1997}. More generally, a substantial magnetic response could be of high interest for many photonic applications, such as the realization of photonic topological insulators \cite{Haldane08, Kanikaev} which so far are limited to the microwave range. An interesting strategy to enhance the response of a laser, and more generally of photonic devices to a magnetic field is to couple photons to matter excitations, sensitive to magnetic fields. One of the most suitable systems to achieve this are exciton-polaritons in semiconductor microcavities. Polariton condensates preserve the intensity and coherence properties of optically pumped VCSELs while, in addition, present both significant magnetic response\cite{Armitage97,Tignon1995, Solnyshkov08, Rahimi2011} and interactions not only among themselves but also with the excitonic reservoir that feeds them\cite{Ferrier2011}. The strong light-matter coupling renders these interactions highly spin-anisotropic\cite{Shelykh10,Solnyshkov08}. All these properties provide a rich playground to manipulate polariton condensates and their polarisation with a magnetic field.

Up to now, the spin degree of freedom in polariton condensates under non-resonant excitation has been mainly considered under the assumption of thermodynamic equilibrium between the spin species and negligible condensate depletion. In the absence of an external magnetic field, one of the consequences of spin anisotropic interactions is that a polariton condensate must be linearly polarized~\cite{Shelykh06, Kasprzak06, Kasprzak2007} with an energy gap between a circularly and a linearly polarized condensate given by: $\left(\alpha_1-\alpha_2\right)n_c/2$. Here ${n_c}$ is the condensate density and $\alpha_1$ and $\alpha_2$ are  respectively the parallel- and antiparallel-spin interaction constants for polaritons. Under magnetic field, it has been predicted that the spin-anisotropic interactions should screen the polariton Zeeman splitting (ZS)  arising from their excitonic component~\cite{Rubo2006}. This phenomenon is frequently refereed as spin Meissner effect due to the analogy with the physics of superconductors. As a result, the chemical potential does not depend on the field up to a critical value $B_c$, and the polarisation of the emission evolves from linear at zero field to circular at $B_c$. For fields larger than ${B_c}$, a Zeeman-like splitting should reappear.

Despite of the intensive search for the observation of such phenomenon, no conclusive experimental evidence has been provided. Polariton condensation under an applied magnetic field oriented along the growth axis has been reported either under resonant excitation \cite{Walker2011}, where the thermal equilibrium condition cannot be fulfilled directly, or at very negative detuning~\cite{Larionov2010,Fischer2014}, where the characteristics of the system are close to those of VCSELs in the weak coupling regime~\cite{vanDoorn1997}. Additionally, the observed ZSs are smaller than the linewidth and thus difficult to measure precisely. In all cases, lasing is observed not in the ground state but in excited spin states, indicating that the system is out of thermal equilibrium. In such a regime, the steady state is defined by the kinetic equilibrium between pump, phonon-induced relaxation, polariton-polariton scattering, and losses~\cite{Kasprzak08}, which contradicts one of the basic requirements of the spin Meissner effect: {\it quasi-}thermodynamic equilibrium.

In this Letter, we report on polarization resolved emission measurements performed at high magnetic field on a polariton condensate in a GaAs-based semiconductor microcavities. We observe a non-monotonous dependence of the ZS and polarisation as a function of the excitation density.  We show that the effects of the magnetic field  are dominated by mechanisms neglected in previous works ~\cite{Larionov2010,Fischer2014}, namely the condensate depletion and spin-dependent interactions of polaritons with the excitonic reservoir. These effects combined with the slow spin relaxation previously considered ~\cite{Soln07,Vishnevsky12,Fischer2014} pushes the two spin subsystems out of equilibrium the one with respect to the other. While our results evidence the difficulties posed by the observation of the spin-Meissner effect in this type of microcavities, they provide an interesting scheme to manipulate the spin (polarisation) of the coherent light emitted from a microcavity.

\begin{figure}
  \centering
  \includegraphics[width=\columnwidth]{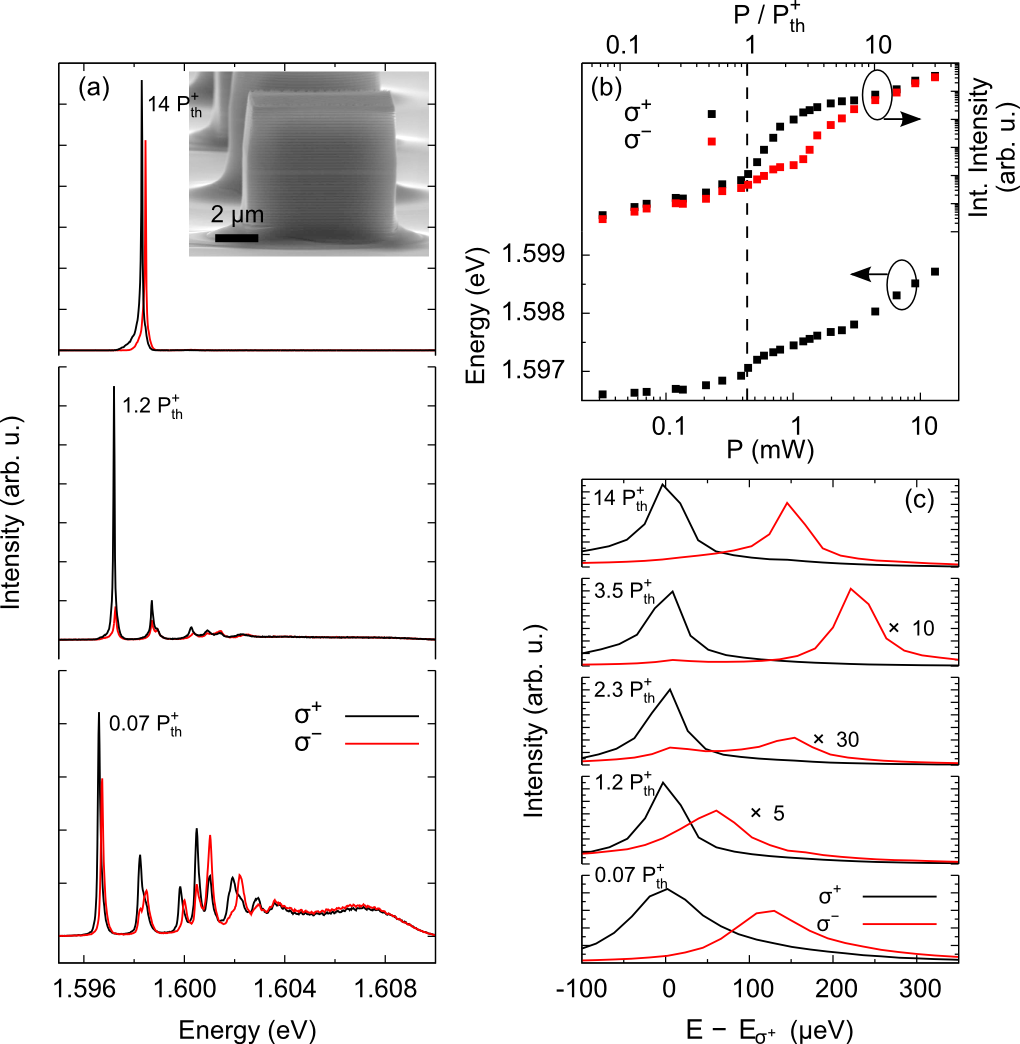}
  \caption{(a) Emission spectra measured under circular polarization for three different excitation powers. $P_\text{th}^+=0.44\, \mathrm{mW}$. (Inset) Scanning electron microscopy image of a micropillar cavity; (b) (top) Integrated intensity for both circular polarizations and (bottom) energy of the $\sigma^+$ polariton mode as a function of excitation power; (c). Emission spectra measured under circular polarization for several excitation powers shown in smaller energy scale. For each power, the zero-energy correspond to the $\sigma^+$ polariton energy. $B=9\,\mathrm{T}$}
  \label{fig:condensation}
\end{figure}

Our sample, described in detail in Ref.~\onlinecite{Bajoni08}, is a $\lambda/2$ cavity with a quality factor exceeding $16 000$ containing 12 GaAs quantum wells (Rabi splitting $\Omega=15\mathrm{meV})$. The micropillar under investigation has a square shape with $3.2\,\upmu\mathrm{m}$ sides (Fig.~\ref{fig:condensation}a). It was designed using electron beam lithography and dry etching.  The exciton-photon detuning used in the experiments was -2.4 meV which corresponds to an excitonic content $x$ larger than 40 \% \cite{Microcavities2007}. The sample was mounted in a magneto-optics cryostat at $T = 1.8\,\mathrm{K}$ and a magnetic field up to $B = 9\,\mathrm{T}$ was applied in the Faraday configuration. The non-resonant excitation was provided by a monomode cw Ti:Sapphire laser tuned typically $100\,\mathrm{meV}$ above the polariton resonances, linearly polarized and focused onto the sample with a spot size around $2\,\upmu\mathrm{m}$.  The emission was collected in a confocal configuration and sent to a CCD camera coupled to a double monochromator. The polariton spin degree of freedom was probed by looking at the circular polarization of the emission~\cite{Shelykh10}.

Figure~\ref{fig:condensation}a shows  the polarisation resolved emission spectra measured at $B= 9\,\mathrm{T}$ for several excitation powers. For the lowest power, several narrow emission lines are observed, each of them corresponding to a well-defined discrete polariton state, confined within the micropillar. As the excitation power is increased, a strong non-linear rise of the emission is observed, first $\sigma^+$ polarised (above $P_{th}^+= 0.44\,\mathrm{mW}$ ), and then for the $\sigma^-$ component (above $ P_{th}^- = 1.06\,\mathrm{mW}=2.4 P_{th}^+$ ). For each circular polarization, the emission is dominated by the lowest energy polariton mode. Importantly, as seen in Fig.~\ref{fig:condensation}b the emission energy undergoes a continuous blueshift when increasing the excitation power, without discontinuity at threshold: this is a key feature allowing to conclude on polariton condensation.

The first hints of the peculiar behaviour of polariton lasing under magnetic field can be seen in Fig.~\ref{fig:condensation}c  which presents the emission spectra within a narrower energy range both for $\sigma^+$ and $\sigma^-$ polarizations. For each excitation power, the ZS is clearly resolved, as it is significantly larger than the polariton linewidth. Strong variations of the ZS are observed when crossing the two condensation thresholds, at $P=P_{th}^+$ and $P = P_{th}^-$.

\begin{figure}
  \centering
  \includegraphics[width=1\columnwidth]{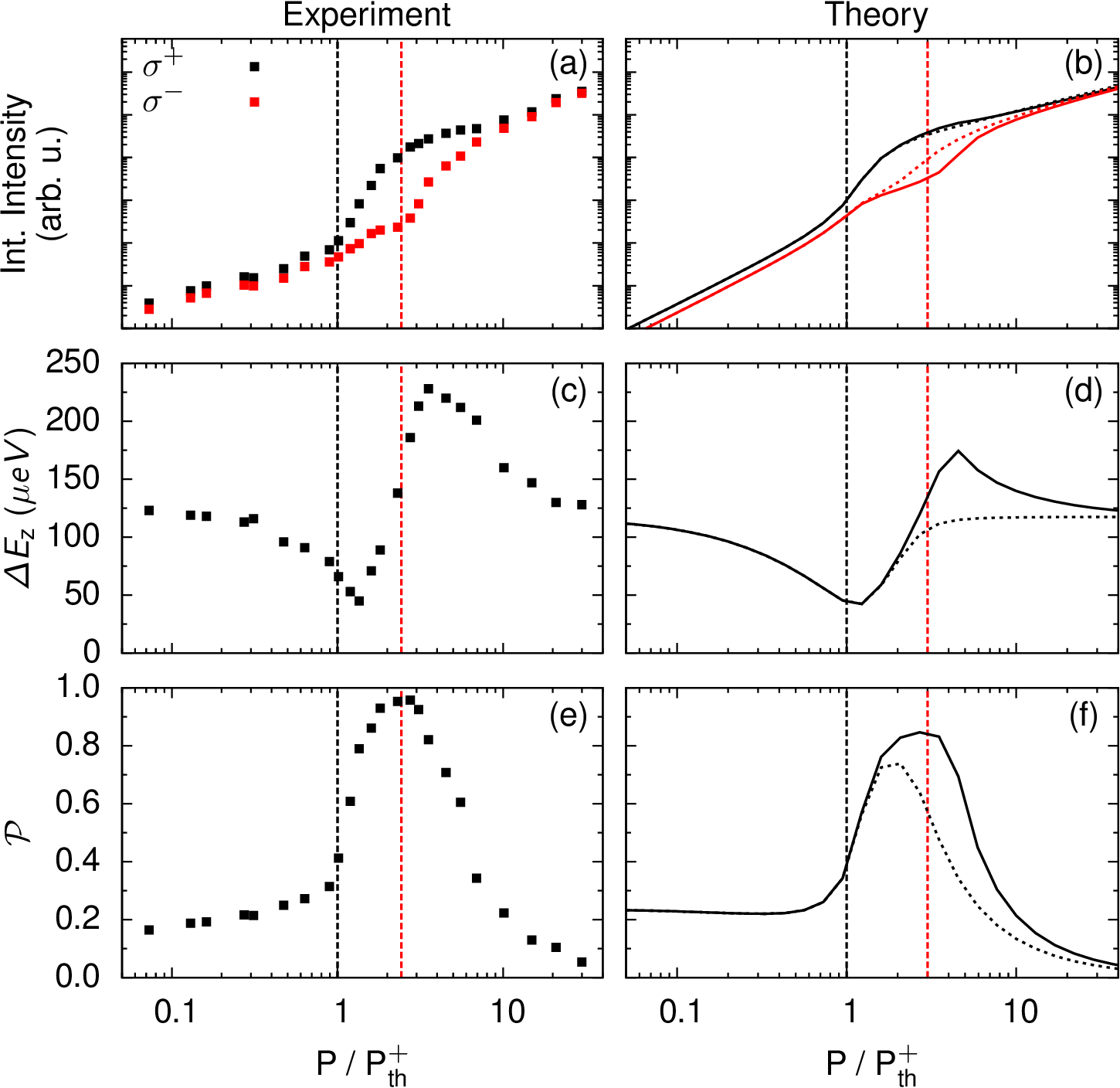}
  \caption{(a) Integrated intensity for both circular polarizations, (c) ZS and (e) circulation polarization ratio measured as a function of power. (b), (d) and (f): (dotted lines) results of the numerical simulation using equations (1-4), and (solid lines) taking into account the nonlinear of the scattering rates described in the text. The vertical dashed line indicate the two thresholds. In the simulation we use: $\tau_\mathrm{ph}= 10 \mathrm{ps}$, $\tau_\mathrm{r} = 300 \mathrm{ps}$, $\alpha_1 =0.6 \mueV$; $\alpha_2 =-0.06 \mueV$, $W_\mathrm{XX}= 10^5\, s^{-1}$ $W_\mathrm{s}=4\times 10^{10}\, s^{-1}$;$\beta =2.2 10^{-3}$;$\delta =2.2 10^{-2}$ .}
  \label{fig:Zeeman_overview}
\end{figure}

The variation of the ZS with excitation density is correlated with the emitted intensity in each component and the degree of circular polarisation, as summarized in Fig.~\ref{fig:Zeeman_overview}. Below $P_{th}^+$, the measured circular polarisation degree is almost constant, of the order of $0.2$ (Fig.~\ref{fig:Zeeman_overview}e). Approaching  $P_{th}^+$, the ZS decreases from $130\,\upmu\mathrm{eV}$ down to a minimum of $45\,\upmu\mathrm{eV}$ (Fig.~\ref{fig:Zeeman_overview}c) which is reached slightly above $P_{th}^+$.
Further increasing the excitation density above $P_{th}^+$, the emission becomes strongly circularly polarized (up to $96 \%$). Surprisingly, and in contradiction with the spin Meissner effect prediction, this is accompanied by a strong increase of the ZS. There is no clear correlation between the circular polarisation degree of the emission and the ZS.
The maximum ZS is reached for $P$ close to $P_{th}^- $ and amounts to $230\,\upmu\mathrm{eV}$, a value significantly larger than the initial one. For larger excitation power, a continuous decay of the ZS is observed down to its initial value, whereas the circular polarization degree goes to zero. The difference $P_{th}^- - P_{th}^+$ between the two condensation threshold powers increases with magnetic field, and the behavior of both polarization ratio and ZS described above, develops progressively, as the magnetic field increases \footnote{See supplementary material for further information}.

In the following, we propose a model to describe these unexpected variations of the ZS across the condensation threshold. We use a kinetic approach, similar to the one presented in Ref.~\onlinecite{Galbiati2012} but including the different spin states and their mutual interaction. We consider the coupled kinetics of four effective states, namely the two spin components of the excitonic reservoir and the lowest energy polariton state. Both exciton reservoirs are assumed to be at the internal thermal equilibrium and their energies are given by \cite{Note1}:

\begin{eqnarray}
\nonumber E_r^{\pm} &=& E_{r_0} \mp 0.5\,{\mu _B}g{B} \\
 &+& {\alpha _1}\left( {n_r^ +  + n_r^ -  + xn_p^{\pm} } \right) + 2{\alpha _2}xn_p^{\mp}
\end{eqnarray}

where $E_{r_0}$ is the reservoir energy at B=0, ${\mu _B}g{B}$ the exciton ZS, ${n_r^{ \pm }}$ the reservoir occupancy for each spin component, ${n_p^{ \pm }}$ the occupation of the polariton state and $x$ their exciton fraction which for simplicity is assumed to be spin independent.  The two excitonic reservoirs are coupled to each other through a spin conversion rate (spin relaxation due, for example, to the Dyakonov-Perel mechanism)\cite{Dyakonov71, Maialle93}. They are also coupled to the polariton state with the same spin and with an energy given by:

\begin{eqnarray}
\nonumber E^{ \pm}& = & E_{r_0}-\Delta \mp 0.5\,x{\mu _B}g{B} + {\alpha _1}x\left( {n_r^{ \pm }  + xn_p^{\pm }} \right) \\
& + & x{\alpha _2}\left( {n_r^{ \mp } + xn_p^ { \mp }} \right)
\end{eqnarray}

where  $E_{r_0}-\Delta$ is the lowest polariton energy at B=0 \cite{Microcavities2007}. In the present experiment $\Delta \simeq 10 meV$ which means that polariton states are about 10 meV away from excitonic states (such as biexcitons for instance). Therefore we do not expect strong variations of  $\alpha_1$ and $\alpha_2$ versus experimental parameters (as recently reported \cite{Vladimirova2010,Deveaud2014}). Therefore in our simulations, $\alpha_1$ and $\alpha_2$ are taken as constant with $\left|\alpha_2\right|$ at least an order of magnitude smaller than $ \left|\alpha_1\right| $.

The time evolution of the state occupations is described by the following set of semi-classical Boltzmann equations:

\begin{align}
\frac{dn_p^{ \pm }}{dt}  = & W_\mathrm{XX}{(n_r^{ \pm })}^2 (n_p^{ \pm } + 1) - W_\mathrm{XX} D n_r^{ \pm } n_p^{ \pm } e^{\frac{E^{ \pm }-E_r^{ \pm }}{k_\mathrm{B}T}} \\ \notag
	& - n_r^{ \pm } / \tau_\mathrm{pol}
	\end{align}
	\begin{eqnarray}
\frac{d  n_r^{ \pm }}{dt}  = W_\mathrm{XX} D n_r^{ \pm} n_p^{ \pm } e^{\frac{E^{ \pm }-E_r^{ \pm }}{k_\mathrm{B}T}} - W_\mathrm{XX}{(n_r^{ \pm })}^2 (n_p^{ \pm }+ 1)
  \\ \notag
 \pm W_\mathrm{s} \frac{e^{\frac{E_{\mathrm{r}}^\pm - E_{\mathrm{r}}^\mp}{k_\mathrm{B}T}}}{1+e^{\frac{E_{\mathrm{r}}^\pm - E_{\mathrm{r}}^\mp}{k_\mathrm{B}T}}} (n_r^{ - }e^{\frac{E_{\mathrm{r}}^\pm - E_{\mathrm{r}}^\mp}{k_\mathrm{B}T}}- n_r^{ + })- n_r^{ \pm } / \tau_\mathrm{r}+P .
\end{eqnarray}

$\tau_\mathrm{pol}$ and $\tau_\mathrm{r}$ are the polariton and reservoir lifetime, $W_\mathrm{XX}$ scattering rate from each spin reservoir into the ground state of the same spin component assisted mostly by exciton-exciton interactions and $D=\frac{Sk_BT}{2\pi \hbar^2} $, $S$ being the pillar surface. The scattering from the reservoir to the condensate is thus assumed to be spin conserving. The spin relaxation rate within the reservoir is given by $W_\mathrm{s}$. We assume that the linearly polarized pump described by the term $P$ introduces equal amounts of ${\sigma ^{ \pm }}$ excitons in the reservoir.
One can notice that there is no spin relaxation term for the polariton state. While polariton condensates often demonstrate coherent spin precession induced by polarisation splitting between linearly polarised eigenstates \cite{Shelykh10}, this splitting is small in the pillar we consider and is neglected.
The time evolution of the system obtained within this model is shown with dashed lines on Fig.~\ref{fig:Zeeman_overview}. Below condensation thresholds, most of the particles are in the reservoir and the polariton population is very small. The total reservoir density grows as $P{\tau _r}$. Since the spin relaxation rate is faster than the scattering into the polariton states (i.e. $W_\mathrm{s}\gg W_\mathrm{XX}n_r$) , the spin populations in the reservoir are close to equilibrium $n_r^+ /n_r^-  \approx \exp \left( \mu_BgB/k_B T \right)$. This finite polarization degree in the reservoir is responsible for the degree of circular polarization of $0.2$ observed in the polariton emission. Such spin transfer from the reservoir to a polariton condensate, due to spin conservation in the energy relaxation of polaritons, is in line with the recent observations of circularly polarized condensates, without magnetic field, but under circular pumping \cite{Kammann12}. Another consequence of the reservoir spin polarization is the renormalization of the polariton ZS. Indeed, below threshold, the ZS is approximately given by $x{\mu _B}g{B_z}+\alpha _1x(n_r^{+}-n_r^{-})$.
At very low density, the polariton ZS is thus equal to  the exciton ZS weighted by the polariton exciton fraction. If $n_r^+$ exceeds $n_r^ -$, the blue shift induced by the reservoir population is larger for $\sigma^+$ (the lowest energy polariton state) than for $\sigma^-$. As a result, as the excitation density is increased, the ZS progressively decreases, as observed experimentally. In this regime, the polariton ZS can even change sign as recently reported \cite{Fischer2014} but with a different physical interpretation.

At some threshold density $n_r^{th}$, the condensation occurs selectively in the more populated component ${\sigma ^ + }$. The scattering between the ${\sigma ^ + }$ reservoir and the ${\sigma ^ + }$ polariton states becomes stimulated, proportional to ${W_{XX}}n_p^ + $, and can thus easily exceed $W_\mathrm{s} $. Above $P_{th}^+$, ${n_r^+}$ does not grow versus pumping, keeping its threshold value, because all extra particles introduced by the pumping are rapidly scattered to the condensate.  ${(n_p^{+}+ n_r^{+})}$ rises approximately as $P{\tau _{pol}}$, $\tau _{pol}$ being typically ten times smaller than the reservoir lifetime. The ${\sigma ^ +}$ condensation therefore leads to a strong reduction of the average time that a ${\sigma ^ +}$ particle spends in the system between absorption and  emission. For $P_{th}^+<P<P_{th}^-$, ${n_r^ - }$ continues to rise as  $P{\tau _r}$, which is thus faster than the increase of the total density of spin up particles given by  ${n_r^+ +n_p^+ }$ . The reservoir population imbalance between the two spin components therefore reduces which induces a rise of the ZS of the polariton mode. Both in the experiment and the theory, slightly above the $P_{th}^+$, the ZS reaches its minimum value and then increases. Interestingly, the condensate circular polarization degree continues to rise well beyond the ZS has reached its minimum. This demonstrates the absence of a simple connection between the circular polarization degree and the ZS, because of the complex role of the reservoir characterized by a finite spin relaxation time.

Increasing the pumping further leads to the condensation of the ${\sigma ^ - }$ spin component. Within the simple set of equations presented above, the reservoir density required to trigger condensation is the same for both spin components. This means that reaching $P=P_{th}^-$,  the total ${\sigma ^ - }$ density is expected to remain smaller than the total ${\sigma ^ + }$ density. The ZS should therefore stay smaller than its low density value as shown by the calculation displayed in solid lines.  When increasing further the pumping, the relaxation rates accelerate and the two spin subsystems become more and more isolated. The same amount of ${\sigma ^ + }$ and ${\sigma ^ - }$ being introduced by the pumping, they are scattered to the condensate and emitted by the condensate before any spin conversion can take place. The ZS therefore goes back to its low density value and the circular polarisation degree of the emission tends to zero. Each spin subsystem can be properly described by a thermodynamic equilibrium, which is not the case of the whole system because of the slow spontaneous spin relaxation rate. This last feature is clearly observed experimentally and reproduced by the above described kinetic approach.

One experimental feature not reproduced by the  model is the ZS pronounced maximum which lies close to the $\sigma^-$ threshold. If we assume that $\alpha_1$ and $\alpha_2$ are constant, the only possible explanation of the ZS maximum is that the $\sigma^-$ total density in the system becomes larger than the $\sigma^+$ density. It means that the $\sigma^-$ threshold is notably increased with respect to the $\sigma^+$ threshold.  We also notice that the  $\sigma^-$ emission shows a plateau just after the $\sigma^+$ threshold, further indicating that the $\sigma^-$ polariton relaxation is affected by the presence of the $\sigma^+$ condensate. This non-linear change of the threshold is not expected from the set of equations (1-4). It implies a non-linear decay of the lifetime and/or of the scattering rates versus the $\sigma^+$ condensate density. A non-linear dependence of the scattering rates over densities can appear if higher order terms are included in the derivation of Boltzmann equations or by taking into account spatial effects. Indeed the polariton scattering rates strongly depend on the overlap integrals between the polariton states and the reservoir. These overlaps can vary significantly in the condensation regime, as shown for instance in the case of focused excitation spots\cite{Ferrier2011,Galbiati2012}. Several physical mechanisms such as reservoir depletion, interaction induced wavefunction change may play a role simultaneously in the evolution of these overlap integrals \cite{Note1}. Their relative contributions are extremely hard to check experimentally. In the first order and following a phenomenological approach, we calculate the polariton dynamics introducing a  non-linear dependence of the scattering rates written as follows:

\begin{equation}
W_\mathrm{XX}^{\mp} = W^0_\mathrm{XX} \left(  1- \frac{ \beta n_p^{\pm}}{1 + \delta  n_p^{\mp}} \right),
\end{equation}

where $\beta$ and $\delta$ are fitting parameters \cite{Note1}.

 The results are shown with solid lines on Fig.~\ref{fig:Zeeman_overview}(b,d,f). The plateau of the $\sigma^-$ emission above the $\sigma^+$ threshold is reproduced together with the increase of the ZS above its low density value. Thus, we infer that the pronounced maximum in the ZS is a signature of a non-linear interplay between the two spin components. Further investigation would be necessary to verify the nature of this non-linear coupling. Note that in the whole density regime explored here, polariton-exciton interactions are much stronger than polariton-polariton interactions. Therefore the ZS variations reflects modulations in the two reservoir densities.

 To conclude, we have measured and described theoretically the polarization degree and ZS of a polariton condensate in a magnetic field in the Faraday configuration. This is of high importance, since the magnetic response of polaritons is a unique opportunity to manipulate photonic modes at optical frequencies using magnetic fields. We report a very strong impact of the spin anisotropic interactions on the condensation dynamics in the magnetic field. However, there is no pumping range in which the original spin-Meissner effect could be observed. This work shows that thermodynamic approaches, which give satisfactory descriptions in some limiting scalar cases, is not fully adapted to describe a spin polarized driven-dissipative condensate, essentially because of the long spin conversion rates. This can be seen as an opportunity to manipulate the spin of polariton condensates using either magnetic fields as in the present work or reservoirs pumped optically with circularly polarised light~\cite{Kammann12}.

We thank A. Kavokin for fruitful discussions. This work was partly supported by the FP7 ITN "Clermont4" (235114),
 by the ANR project Quandyde (ANR-11-BS10-001), by the French RENATECH network, by the Labex NanoSaclay, and the RTRA Triangle de la Physique contract "Boseflow1D" and "2012-039T-InterPol".


%

 \newpage
 

\section*{Supplementary material}

# Supplementary Material for

## "Driven-dissipative confined polariton condensate under magnetic field"

C. Sturm[1], D. Solnyshkov[2], O. Krebs[1], A. Lemaître[1], I. Sagnes[1], E. Galopin[1], A. Amo[1], G. Malpuech[2], and J. Bloch[1]

[1] Laboratoire de Photonique et Nanostructures, LPN/CNRS, Route de Nozay, 91460 Marcoussis, France

[2] Institut Pascal, Photon-N2, Clermont Université and Université Blaise Pascal, CNRS, 63177 Aubière Cedex, France

# I. Additional measurements for different values of the magnetic field:

We present here measurements for three values of the magnetic field, showing that all the observed features described in the text (variation of the Zeeman splitting and of the circular polarization across the condensation thresholds) progressively develop as the magnetic field increases.

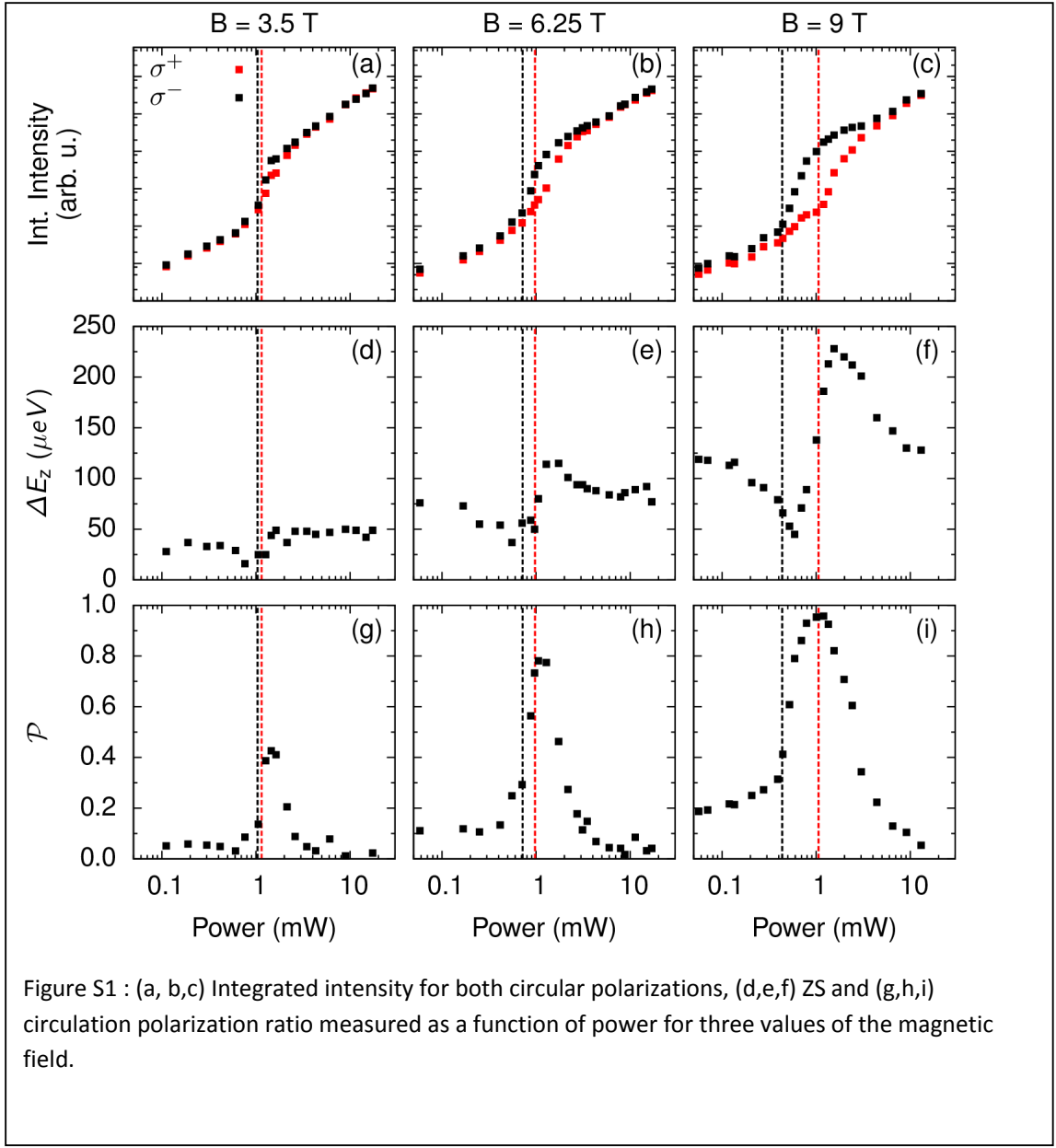

Figure S1 : (a, b,c) Integrated intensity for both circular polarizations, (d,e,f) ZS and (g,h,i) circulation polarization ratio measured as a function of power for three values of the magnetic field.

## II. Interaction constants

We describe below in more detail the assumptions used in the model concerning the different interaction constants.

We assume that the interactions between 2 exciton-polaritons with same spins are described by $x_1 x_2 \alpha_1$ where $x_{1,2}$ are the exciton fractions of the two interacting exciton-polaritons, and $\alpha_1$ is the interaction constant for two excitons having the same spin.

We assume that the interaction constant between two reservoir excitons having opposite spins is also equal to $\alpha_1$. Therefore, the interactions in the reservoir are assumed to be spin-isotropic.

However, the interaction between exciton-polaritons with opposite spins is a second order process which passes through an intermediate state: either dark excitons or biexcitons. It approximately reads [Ref. 4 of the main text]: $\alpha_2 \approx |M|^2 / \Delta E$, where $\Delta E$ is the difference between the energies of the interacting particles and of the intermediate state. If we consider the interaction between two polaritons, $\Delta E \sim -\Omega_R$ where $\Omega_R$ is the Rabi splitting. If we consider the interaction between a polariton and a reservoir exciton, the energy difference is $\Delta E \sim -\Omega_R / 2$. The resulting interaction constant is therefore about twice larger. In the main text, we label $x^2 \alpha_2$ the interaction constant between two polaritons, and $2x\alpha_2$ the interaction constant between a polariton and a reservoir exciton. However, one should notice that in our simulation the interaction between opposite spin particles always play negligible role because of the low value of $\alpha_2$ and the low value of the condensate density with respect to the reservoir density close in the explored range of excitation powers.

## III. Spatial effects modifying polariton relaxation

We discuss here two spatial non-linear mechanisms which could be responsible for the non-linear change of the scattering rates as described in the equation (5) of the main text. In both cases, our goal is to construct a first-order correction (linear in $n_p$) of the scattering rates due to the spatial effects. This correction should saturate when $n_p$ becomes large.

**First possible mechanism:** An important mechanism is the drain of particles from the reservoir to the $\sigma^+$ condensate. The drain depletes the reservoir in a stronger way at the center of the pillar than on the edges. This reservoir minimum is located exactly where the polariton wavefunction is maximal. It tends to reduce the effective scattering rate between the reservoir and the polariton state. Because of the spin conversion in the reservoir, the hole in the density also appears in the $\sigma^-$ reservoir, inducing a decrease of the scattering rate, which in turn increases the $\sigma^-$ condensation threshold. The $\sigma^-$ reservoir therefore becomes more populated than the $\sigma^+$ one, which eventually leads to an increase of the Zeeman splitting above its low density value as observed experimentally.

Let us first describe this spatial effect within a single spin component, to simplify the picture. We can expect, that at large $n_p$, not all particles of the reservoir will be able to effectively scatter towards the condensate, because they will be mostly located in the areas where the condensate density is small (wavefunction tails). Let $f$ be the fraction of the reservoir accessible in this limit:

$$W_{XX} \xrightarrow[n_p \to \infty]{} f W_{XX} \qquad (1)$$

Then, in order to provide this saturation and to remain linear in $n_p$, the correction has to be written as ( we consider a single spin component for simplicity):

$$W_{XX} = W_{XX}^0 \left(1 - (1-f)\frac{n_p}{c+n_p}\right) \qquad (2)$$

where $c$ is a constant, whose dimensionality is that of the number of particles. $c$ should be proportional to the reservoir density at threshold $c = \xi n_{r0}$, and its value describes the efficiency of the mechanism at low $n_p$ ($n_p \ll c$):

$$W_{XX} \approx W_{XX}^0 \left(1 - \frac{(1-f)n_p}{\xi n_{r0}}\right) \qquad (3)$$

We can finally reformulate our correction in order to minimize the number of fitting parameters:

$$W_{XX} = W_{XX}^0 \left(1 - \frac{\beta n_p}{1 + \delta n_p}\right) \qquad (4)$$

Where $\beta = (1-f)/\xi n_{r0}$, $\delta = 1/\xi n_{r0}$ are the two fitting parameters which have a clear physical meaning. This effect can take place within a single spin component as written above. However, because of the spin conversion in the reservoir, the depletion of one spin component also induces depletion for the other spin component in the reservoir. The corresponding correction of the scattering rate reads :

$$W_{XX}^{\mp} = W_{XX}^0 \left(1 - \frac{\beta n_p^{\pm}}{1 + \delta n_p^{\mp}}\right) \qquad (5)$$

This formula is nothing else but the equation (5) of the main text. In fact, it is the only correction which is relevant for the present study. It is more important both in its values and its consequences, than the correction within a single spin component. Indeed, a small decrease of the scattering rate of the $\sigma^+$ component above its threshold is masked by a nonlinear increase of the relaxation due to the stimulation. On the other hand, when the $\sigma^-$ component is not yet condensed, its dynamics is much more sensitive to a decrease of its scattering rate. The maximum value that can reach the intra-spin correction is given by $\beta/\delta = (1-f)$. It is the limit of the formula 4 when $n_p$ goes to infinity. This maximum correction is found to be relatively small from our fitting, of the order of 0.1.

On the other hand, the inter-spin correction can reach values larger than $\beta/\delta$, because the value of the minority component can remain small, while the other component is already large.

The equation (5) of the main text only accounts for inter-spin corrections. The relative importance of various contributions is illustrated on the figure S2. We plot the dependence on the pumping of the scattering rates (normalized by the low-density value $W_{XX}^0$) for the two spin components induced by the above mentioned mechanism. The parameters are the ones used to fit the experimental data (figure 3-b,d,f of the main text). The values of the fitting parameters used were: $\beta = 2.2 \times 10^{-3}$, $\delta = 2.2 \times 10^{-2}$. The figure S2 shows the evolution of the total scattering rates (solid lines) and the contributions of the mechanisms within a single component (dotted lines) and between the spin components (dashed lines). We see that the most important effect is due to the action of the majority component on the minority one (red dashed line). Since the variation of the $\sigma^-$ scattering rate occurs below threshold, it increases the threshold value for $\sigma^-$ polaritons and, consequently, the measured Zeeman splitting. Both curves converge towards a constant value corresponding to the high-density limit.

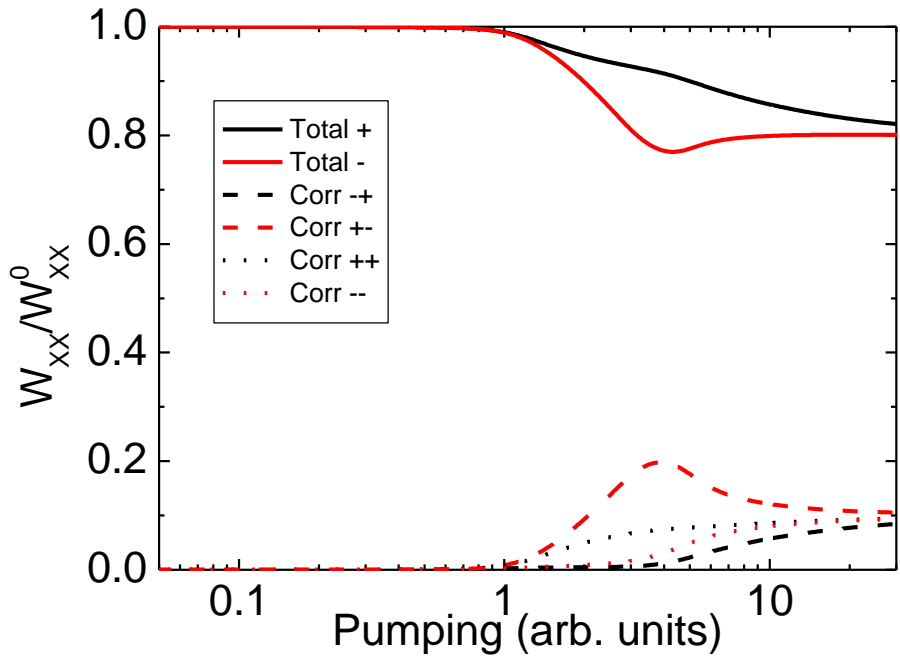

Figure S2. Modification of the scattering rate possibly due to spatial effects, as described by Eq. (5) for the magnetic field of 9 T (right column of fig. S2). Black - $\sigma^+$ polaritons, red - $\sigma^-$ polaritons. Solid lines show the evolution of the total scattering rates, dotted lines – the contribution within the same component, dashed lines – between the two components.

**Another mechanism** is based on the spin-anisotropic polariton-polariton interactions. The central hypothesis is that a $\sigma^+$ condensate with approximately a cosine wavefunction forms at the center of

the pillar. This condensate interacts attractively with the $\sigma^-$ polaritons through the $\alpha_2$ constant. The $\sigma^-$ polariton wave function is therefore expected to be more localized in real space than in the absence of the $\sigma^+$ condensate, which reduces its overlap integral with the reservoir which is almost homogeneous, with a small depletion at the center of the pillar as mentioned above. Again this decrease of the overlap integral reduces the scattering rate involving the $\sigma^-$ particles, once the $\sigma^+$ condensate becomes strongly occupied.

In this case, we can assume that the scattering rate is reduced proportionally to the relative width of the condensate wavefunction relative to the size of the reservoir (size of the pillar). The relative shrinking of the $\sigma^-$ wavefunction due to the interactions with $\sigma^+$ is calculated using the perturbation theory. At low $n_p^+$, it is proportional to $\alpha_2 n_p^+ / E_c$, where $E_c$ is the localization energy of the condensate $E_c = \pi^2 \hbar^2 / mL^2$, where $L$ is the pillar size. As the interactions in $\sigma^-$ become important, the efficiency of the scattering in the limiting case is described by the competition between the repulsion $\alpha_1 n_p^-$ within $\sigma^-$ (because the repulsive interactions due to $\alpha_1$ lead to the spreading of the wavefunction towards the TF profile) and contraction due to $\sigma^+$, as before, and therefore the correction in this limit is expressed as a fraction $\alpha_2 n_p^+ / \alpha_1 n_p^-$. Both limits are incorporated in a single formula as:

$$W_{XX}^- = W_{XX}^0 \left(1 + \frac{\alpha_2 n_p^+}{E_c + \alpha_1 n_p^-}\right) = W_{XX}^0 \left(1 - \frac{\beta n_p^+}{1 + \delta n_p^-}\right) \quad (6)$$

Where for this mechanism our fitting parameters are $\beta = -\alpha_2 / E_c$, $\delta = \alpha_1 / E_c$.

Notice that the modification of the reservoir profile can also affect the wavefunction of the polariton condensate. The hole in each component creates a potential minimum for polaritons of the same spin, leading to their localization. On the other hand, it has a delocalizing action on the opposite component through the $\alpha_2$ constant. Again, only the inter-species contribution appears relevant to describe the experimental situation.

The relative importance of all these mechanisms depends of the exact values of the parameters. As written in the main text, it is very hard experimentally to access to the spin dependent reservoir distribution, and to check which of these mechanisms could be dominant. This is why we describe the polariton dynamics using a phenomenological approach summarized by equation 5.

### III. Additional simulations corresponding to figure S1

Below we show the results of the simulations using the same parameters as in fig. 3 of the main text but for the three values of the magnetic field (as in fig. S1). The model gives very good qualitative description of the observed features. The values of the fitting parameters used were: $\beta = 2.2 \times 10^{-3}$, $\delta = 2.2 \times 10^{-2}$.

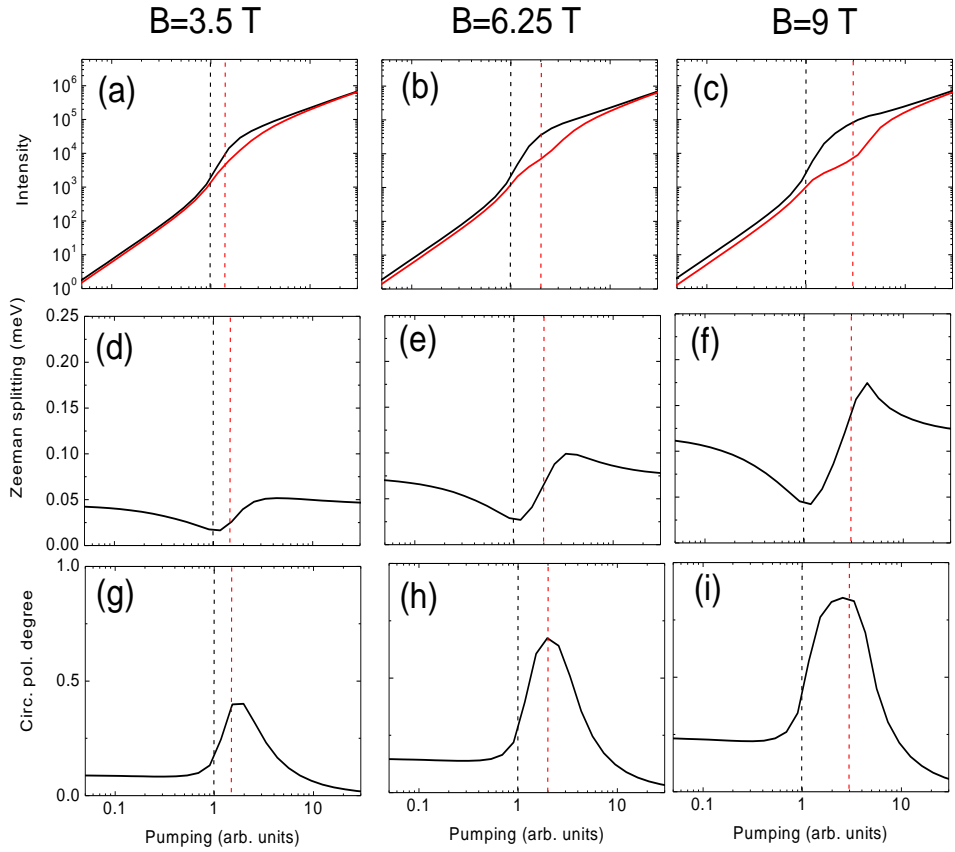

Figure S3. Theoretical simulations of the polariton emission intensities, Zeeman splitting values and the circular polarization degree as a function of pumping for 3 different values of the magnetic field.



\end{document}